\documentclass[epj]{svjour}
\usepackage{graphics}
\begin{document}
\title{Fitting the Power-law Distribution to the 
Mexican Stock Market index data.}

\author{H.F. Coronel-Brizio\  \thanks{\emph{e-mail:} hcoronel@uv.mx},
C.R. de la Cruz-Laso\ \thanks{\emph{e-mail:} cecruz@uv.mx} 
 and\ A.R. Hernandez-Montoya\ \thanks{\emph{e-mail:} alhernandez@uv.mx }%
}

\institute{Facultad de F\'{\i}sica e Inteligencia Artificial.
Universidad Veracruzana, Apdo. Postal 475. Xalapa, Veracruz. M\'{e}xico.}
\date{}

\abstract{
%In this paper, 
In the spirit of the emergent field of econophysics, a goodness-of-fit test
for the Power-Law distribution, based on the Empirical Distribution Function (EDF) is 
presented, and related problems are discussed. An analysis of the tail
behaviour of the daily logarithmic variation of the Mexican Stock Market Index (IPC), showed
distributional properties which are consistent with previous studies. 
 % end of PACS codes
} %end of abstract

\authorrunning{H.F. Coronel-Brizio \it{et al.}}

\titlerunning{Test of fit for the Power-law Distribution and its application to the Mexican stock Market }
%index relative changes }

\maketitle
\section{Introduction}
\label{intro}

The Power-law distribution, also called Pareto distribution describes 
phenomena presented in fields as Social Sciences, i.e. Economics 
and Finance (individuals income distribution, stock market price variation 
distribution, etc) or Physics (phase transitions, nonlinear dynamics, and 
disordered systems). Although with a tradition of more than a hundred years 
\cite{VP97,LB00,roehner}, interest of physics community in the complex behavior of Financial Markets has strongly increased in the last years boosted by the availability of 
huge worldwide economical data electronically recorded, giving rise to a complementary or 
non orthodox under the point of view of traditional Economic Theory way to 
attack problems based in the empirical data analysis rather than in the traditional economical 
analysis. The collection of methods and techniques originally developed to study 
problems arising from physics and that currently are being used to understand 
financial complex systems, is called Econophysics and appears as an emergent branch 
of Physics by itself ~\cite{roehner,Takayasu,fisica1,fisica2,fisica3,mantegna}.
\\
About 100 years ago, the Power-Law distribution was proposed by Pareto to 
describe the distribution of income of individuals. More recently, based in Mandelbrot's 
pioneer work ~\cite{mandelbrot} and later of Mantegna and Stanley~\cite{mantegna1}, analyses of price 
distribution variations of leading stock markets~\cite{GOPI98,GOPI99,liu1,lux} and 
individual companies~\cite{liu2,plerou} have been reported; all of them showing Pareto tails, 
with  $\alpha \simeq 3$ for the stock market case.
\\
In section \ref{Applica} of this work, we analyze the distribution of daily logarithmic  
differences of the Mexican IPC stock index, defined as $S(t):=\log Z(t+1)- \log Z(t)$; 
for IPC values of $Z(t)$ recorded during an almost  10 years period, from April 14, 1990 
until December 31, 2002.
\\
A goodness-of-fit test based on the Empirical Distribution Function (EDF)
is introduced in sections ~\ref{Empiri} to~\ref{asympto}. Section \ref{Testpro} explains the 
test procedures and section \ref{Montec} shows results of a Monte Carlo study used to investigate
the small sample distribution of the test statistics and the speed of convergence to their
asymptotic distribution.
\section{Empirical Distribution Function (EDF).}
\label{Empiri}
%\vspace*{-.09cm}

Let $Y_{1},\ldots,Y_{n}$ be a random sample from an
absolutely continuous distribution $F$ and
suppose that we are interested in testing the null hypothesis that
the sample was drawn from the distribution
\begin{equation}\label{pareto1}
F(y)=1-\left(\frac{\gamma}{y}\right)^{\alpha}
\end{equation}
with support on $0 < \gamma \leq y$ for $\alpha>0.$
\\
The distribution (~\ref{pareto1}) is known as the {\em Power-law} distribution.
A test of fit can be based on measures of discrepancy between the empirical distribution
function $F_{n}$ and the hypothesized distribution $F$. Such test statistics
are referred to as {\em empirical distribution function} based
statistics or simply EDF-statistics. Here we will consider statistics within the
class
\begin{equation}\label{Qn}
Q_{n}=n\int_{-\infty}^{\infty}\left[F_{n}-F\right]^{2}\psi dF
\end{equation}
\\
When $\psi(.)=1$, $Q_{n}$ is known as
the Cram\'{e}r von-Mises $W^{2}$ statistic and, for $\psi(\nu)=\left\{\left[F(\nu)\right]
\left[1-F(\nu)\right]\right\}^{-1}$, it is known as the Anderson-Darling
 $A^{2}$ statistic.
\\
The test are based on the quantities $Z_{i}=F(Y_{i};\gamma,\alpha)$, the
Probability Integral Transformation which, under the null hypothesis
produces observations uniformly distributed on $(0,1)$. To obtain
computational formulas for the test statistics, the 
expression in (~\ref{Qn}) can be written in terms of the observed
discrepancy between the empirical distribution function calculated
from the transformed observations and the uniform cumulative
distribution function; i.e.,
\\
\begin{equation}\label{QnZ}
Q_{n}=n\int_{0}^{1}\left[F^{*}_{n}(z)-z\right]^{2}\psi^{\prime}(z)dz
\end{equation}
\\
where $\psi^{\prime}(z)=1$ for $W^{2}$ and $\psi^{\prime}(z)=\left[z(1-z)
\right]^{-1}$ for the Anderson-Darling statistic $A^{2}$.
\\
Computational formulas for these statistics involve the ordered sample
values $Z_{(1)}< \ldots Z_{(n)}$:

\begin{eqnarray}
W^{2}&=&\sum_{i=1}^{n}\left[Z_{(i)}-\left(2i-1\right)/(2n)\right]^{2}
+1/(12n)  \label{w2} 
\end{eqnarray}

\begin{center}
\begin{equation}
A^{2}=-n-(1/n)\sum_{i=1}^{n}(2i-1)\left[\log Z_{(i)}+
\log \left\{1-Z_{(n-i+1)}\right\} \right]
\label{a2}
\end{equation}
\end{center}

\section{Estimation}
\label{Maxi}

Given observed values $y_{1},\ldots,y_{n}$ of a random sample from the
distribution (~\ref{pareto1}), the log-likelihood is
\begin{equation} \label{lik}
\lambda(\alpha,\gamma)=n\log \alpha+n\alpha\log \gamma-(\alpha+1)
\sum_{i=1}^{n} \log y_{i}
\end{equation}
\\
When $\gamma$ is known, the maximum-likelihood estimator of
the parameter $\alpha$ is 
\begin{equation} \label{mlealpha}
\hat{\alpha}=\left[\frac{1}{n}\sum_{i=1}^{n} \log \left(y_{i}/ \gamma \right)\right]^{-1}
\end{equation}
\\
If $\gamma$ is unknown, we use an approach proposed by \cite{LS92}
substituting (~\ref{mlealpha}) into (~\ref{lik}) to obtain
the {\em profile log-likelihood} for $\gamma$; namely 

\begin{equation} \label{profile}
\lambda^{*}(\gamma)=n\ln (n)-n\ln \left[-n\ln (\gamma)+\sum _{i=1}^{n}\ln y_{i}\right]-n-\sum _{i=1}^{n}\ln y_{i}
\end{equation}
\\
It can be seen that the above function is increasing in the range
$0<\gamma<\exp(\sum_{i=1}^{n}\log y_{i} /n)$. In fact, since
$d\lambda^{*}(\gamma)/d\gamma=n\hat{\alpha}/ \gamma$, the
derivative is positive in it's range of admissible values. Thus, the 
maximum-profile-likelihood estimator for $\gamma$ is $\hat{\gamma}=y_{(1)}$, the minimum sample value.

When $\gamma_{i}=\gamma \; , i=1,\ldots,n$, is unknown, the estimate $\hat{\gamma}=Y_{(1)}$ is
super-efficient in the sense that it's variance tends to zero faster than $1/n$. Using
this estimate, we "loose" one sample observation and there are computational problems for calculating
$A^{2}$. Since, in this case, $z_{(1)}=0$, we suggest to estimate the parameter $\gamma$ by finding
the value $\tilde{\gamma}$, of $\gamma$, which satisfies
\begin{equation} \label{mleh}
\gamma-y_{(1)}\left[1-\left\{n\alpha(\gamma)\right\}^{-1}\right]=0
\end{equation}
where $\alpha(\gamma)$ is defined by (~\ref{mlealpha}). Thus, starting with $\gamma=y_{(1)}$, we
search for the solution over the interval $(0,y_{(1)})$. This method of estimation does
not seem to have a significant effect over the sampling distributions of the test statistics
as it will be indicated from the results of the section ~\ref{Montec}.

\section{ Asymptotic distributions }
\label{asympto}

EDF asymptotic distribution statistics was obtained applying the theory in {\cite{DURBIN73}. The process $\sqrt{n}\left\{F_{n}(x)-F(x)\right\}$ evaluated at $t=F(x)$, converges
weakly to $\left\{Y(t),t \in (0,1) \right\}$, a Gaussian process with
zero mean and covariance function $\rho(s,t)$ which depends on the
parameters estimated and $F$. The statistics $W^2$ and 
$A^2$ are asymptotically functions of $Y(t)$; namely, 
$W^{2} \stackrel{{\cal D}}{\longrightarrow}\int_{0}^{1} Y^{2}(t)dt$, 
$A^{2} \stackrel{{\cal D}}{\longrightarrow}\int_{0}^{1} a^{2}(t)dt$ ,
where $a(t)=\frac{Y(t)}{\sqrt{t(1-t)}}$.
Let $\rho^{*}(s,t)$ denote
the covariance function for a given statistic. The limiting distribution
of the test statistic is that of $\sum^{\infty} \lambda_{j} \nu_{j}$, where
$\nu_{1},\nu_{2},\ldots$ are independent $\chi^{2}_{(1)}$ random variables
and $\lambda_{1},\lambda_{2},\ldots$ are the eigenvalues of the integral
equation
\begin{equation}
\int_{0}^{1} \rho^{*}(s,t)f_{j}(s)ds=\lambda_{j}f_{j}(t)
\label{eqint}
\end{equation}
\\
When the parameters are known, the covariance function is given by 
$\rho(s,t)=min(s,t)-st$. When $\gamma$ is known, and $\alpha$ is estimated using (~\ref{mlealpha}), the covariance
function for the limiting process becomes $\rho^{*}(s,t)=\rho(s,t)-(1-s)(1-t)\log(1-s)\log(1-t)$
which corresponds to the same covariance function as the resulting one when testing fit to the
exponential distribution for unknown $\alpha$, so the asymptotic
distribution of the test statistics is the same for testing exponentially or
fit to the Power-law distribution with $\alpha$ unknown. Tables 4.2 and 4.11 in 
\cite{DS86}, give selected percentage points for the case of known parameters,
and for testing exponentially with unknown $\alpha$, respectively.
They are summarized in table ~\ref{table1} for quick
reference.
\begin{table}
\caption{Asymptotic percentage points of $W^2$ and $A^{2}$ statistics}
\label{table1}
\begin{tabular}{lllllll}
%\begin{tabular}{||r|r|r|r|r|r|r||}
\hline\noalign{\smallskip}
%\multicolumn{1}{r}{ } & \multicolumn{6}{c}{Significance level} \\
%\hline
%Both parameters known &0.250&0.15&0.10&0.05&0.025&0.010\\
Both  &&&&&&\\
Parameters &0.250&0.15&0.10&0.05&0.025&0.010\\
known &&&&&&\\
\noalign{\smallskip}\hline\noalign{\smallskip}
%\hline
%\hline
 $W^{2}$&0.209&0.284&0.347&0.461&0.581&0.743 \\
 $A^{2}$&1.248&1.610&1.933&2.492&3.070&3.880 \\
%\hline
\hline
$\alpha$ unknown&0.25&0.15&0.10&0.05&0.025&0.010\\
\hline
%\hline
$W^{2}$&0.116&0.148&0.175&0.222&0.271&0.338 \\
 $A^{2}$&0.736&0.916&1.062&1.321&1.591&1.959 \\
%\hline
\noalign{\smallskip}\hline
\end{tabular}
\end{table}
\section{Test procedures}
\label{Testpro}

\begin{enumerate}
\item Given the ordered sample values $y_{(1)}\leq \ldots \leq y_{(n)}$, the test statistics are
computed from the values $z_{(i)}=F(y_{(i)};\alpha,\gamma)$ where the values of the parameters
$\alpha$ and/or $\gamma$ can be replaced by their estimates if they are not known.
\item Compute the value of the test statistic $A^{2}$ using (~\ref{a2}) or $W^{2}$ from (~\ref{w2}).
\item Refer to table ~\ref{table1} for the appropriate case and significance
level.
\item If the value of the test statistic exceeds that from the table, reject the Power-law
model at the corresponding significance level.
\end{enumerate}

\noindent
When $\gamma$ is known, the distribution of $X=\log(y/\gamma)$ is exponential with parameter $\alpha$;
i.e., $F_{X}(x)=1-exp(-\alpha x)$, and the problem of testing fit to the Power-law distribution is
equivalent to that of testing the null hypothesis that the transformed observations $x_{1},\ldots,x_{n}$
were drawn from an exponential distribution. If $Y_{1},\ldots,Y_{n}$ is a sample of $n$ independent
values from Power-law distributions with parameters $\gamma_{1},\ldots,\gamma_{n}$ and the same
value of the parameter $\alpha$, a test of fit can be carried out by transforming $X_{i}=\log(Y_{i}/\gamma_{i})$
and, again, test that the transformed values come from an exponential population. It is important to
remark that this procedure is computationally equivalent to the procedure described in the
previous paragraph, due to the fact that $F_{X}(x_{i})=F_{Y}(y_{i})$. 
\\

\section{Monte Carlo study}
\label{Montec}

\noindent
A simulation study was conducted to investigate the speed of convergence of the empirical
percentage points of the test statistics to their asymptotic values. five thousand samples
of size $n=20(200)20$ were simulated. In the simulation, values of $\gamma$ and $\alpha$ ranging from
$1/4$ to $10$ were used to verify that there was no effect of the parameter values on the speed of
convergence; the results showed little or no effect at all. The empirical percentage points
presented in tables 2 and 3 show typical results obtained for $\gamma=1$ and $\alpha=5$. 

%
%
%
%\begin{table}[htbp]
\begin{table}
\centering
\caption{Empirical percentage points of $W^2$ and $A^{2}$: $\gamma$ known}
\label{table2}
\begin{tabular}{lllllll}
%\begin{tabular}{||r|r|r|r|r|r|r||}
\hline\noalign{\smallskip}
\multicolumn{1}{c}{$W^{2}$} & \multicolumn{6}{c}{Significance level} \\
\hline
%\hline\noalign{\smallskip}
$n$ &0.250&0.15&0.10&0.05&0.025&0.010\\
%\hline\noalign{\smallskip}
\hline
%\hline
 20& 0.116 & 0.147 & 0.174 & 0.223 & 0.270 & 0.328 \\
 40& 0.117 & 0.151 & 0.177 & 0.223 & 0.266 & 0.324 \\
 60& 0.114 & 0.147 & 0.174 & 0.220 & 0.271 & 0.337 \\
 80& 0.116 & 0.147 & 0.173 & 0.219 & 0.272 & 0.335 \\
100& 0.116 & 0.148 & 0.174 & 0.222 & 0.271 &0.336 \\
$\infty$&0.116&0.148&0.175&0.222&0.271&0.338 \\
%\hline
\hline
\multicolumn{1}{c}{$A^{2}$} & \multicolumn{6}{c}{Significance level} \\
\hline
$n$ &0.250&0.15&0.10&0.05&0.025&0.010\\
%\hline
\hline
 20&  .728 & .900 &1.046 &1.302 &1.565 &1.893 \\
 40&  .725 & .903 &1.058 &1.316 &1.587 &1.918 \\
 60&  .737 & .916 &1.054 &1.313 &1.604 &1.940 \\
 80&  .727 & .899 &1.042 &1.274 &1.588 &1.944 \\
100&  .737 & .920 &1.066 &1.305 &1.552 &1.877 \\
$\infty$&0.736&0.916&1.062&1.321&1.591&1.959 \\
%\hline
\hline\noalign{\smallskip}
\end{tabular}
\end{table}
\noindent
Table 2 shows the empirical percentage points for the test-statistics using the known value
of $\gamma$ and estimating $\alpha$ using (~\ref{mlealpha}). These results clearly suggest
that the asymptotic percentage points can be used even for small values of $n$. When 
the parameter $\gamma$ is estimated using (~\ref{mleh}), the speed of convergence of the
empirical percentage points, to those corresponding to the asymptotic distribution, is
slower. As it is shown in table 3, it is recommended the use the
Monte Carlo percentage points for $n \leq 100$.
\\
It is worth to note that the results in table 3 are very close to those reported in table 4.15
from \cite{DS86} for testing exponentially with origin and scale parameters
unknown, which is referred as an external check of these results.

\begin{table}
\centering
\caption{Empirical percentage points of $W^{2}$ and $A^{2}$: $\gamma$ estimated using (~\ref{mleh})}
\label{table3}
\begin{tabular}{lllllll}
%\begin{tabular}{||r|r|r|r|r|r|r||}
\hline\noalign{\smallskip}
\multicolumn{1}{c}{$W^{2}$} & \multicolumn{6}{c}{Significance level} \\
\hline
$n$  &0.250&0.15&0.10&0.05&0.025&0.010\\
\hline
%\hline
%
%
 20& .106&  .133&  .157&  .198&  .241&  .298 \\ 
 40& .112&  .145&  .171&  .212&  .258&  .318 \\ 
 60& .112&  .143&  .172&  .219&  .269&  .325 \\ 
 80& .114&  .145&  .170&  .214&  .276&  .335 \\ 
100& .114&  .147&  .176&  .224&  .273&  .319 \\ 
$\infty$&0.116&0.148&0.175&0.222&0.271&0.338 \\
%\hline
\hline
\multicolumn{1}{c}{$A^{2}$} & \multicolumn{6}{c}{Significance level} \\
\hline\noalign{\smallskip}
%\hline
$n$  &0.250&0.15&0.10&0.05&0.025&0.010\\
\noalign{\smallskip}\hline\noalign{\smallskip}
 20&  0.615&  0.762&  0.880& 1.091& 1.321& 1.602 \\
 40&  0.662&  0.825&  0.955& 1.192& 1.428& 1.710 \\
 60&  0.675&  0.842&  0.973& 1.216& 1.498& 1.826 \\ 
 80&  0.685&  0.862&  0.994& 1.236& 1.502& 1.856 \\ 
100&  0.693&  0.866&  1.005& 1.240& 1.523& 1.814 \\
$\infty$&0.736&0.916&1.062&1.321&1.591&1.959 \\
\hline\noalign{\smallskip}\hline
\end{tabular}
\end{table}
\section{Application to the analysis of stock price variations}
\label{Applica}

The data analyzed consist of two data sets; the first corresponding
to the 126 largest differences $d_{i}$, and the second one corresponding
to the absolute values of the 144 smallest differences $d^{*}_{i}$, from the series
$S(t)=\log Z(t+1)- \log Z(t)$; where $Z(t)$ denotes daily values
of the Mexican IPC stock index, between the April 14, 1990 and December 31, 2002.\\

\begin{figure}
\resizebox{0.5\textwidth}{!}{%
\includegraphics{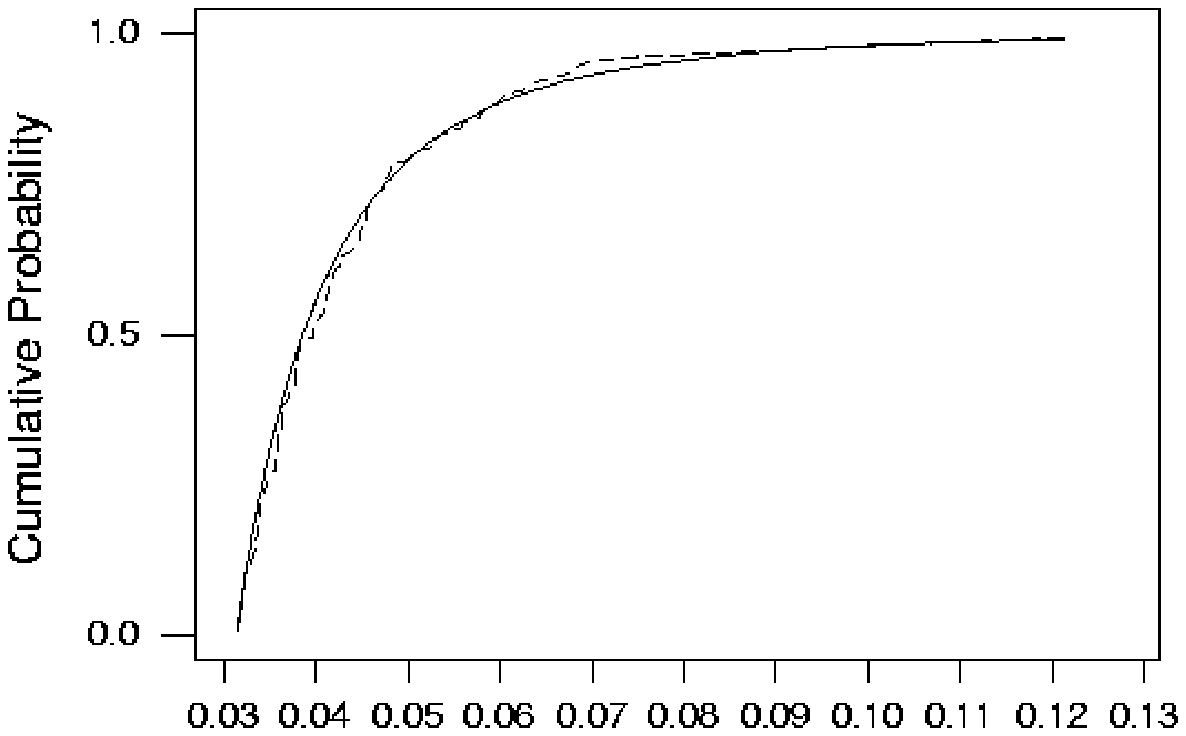}}
\caption{Empirical (dash) and Fitted (solid) CDF for differences $d_{i}$}
\label{plot1}
\end{figure}
\vspace*{-.454cm}
\begin{figure}
\resizebox{0.5\textwidth}{!}{%
\includegraphics{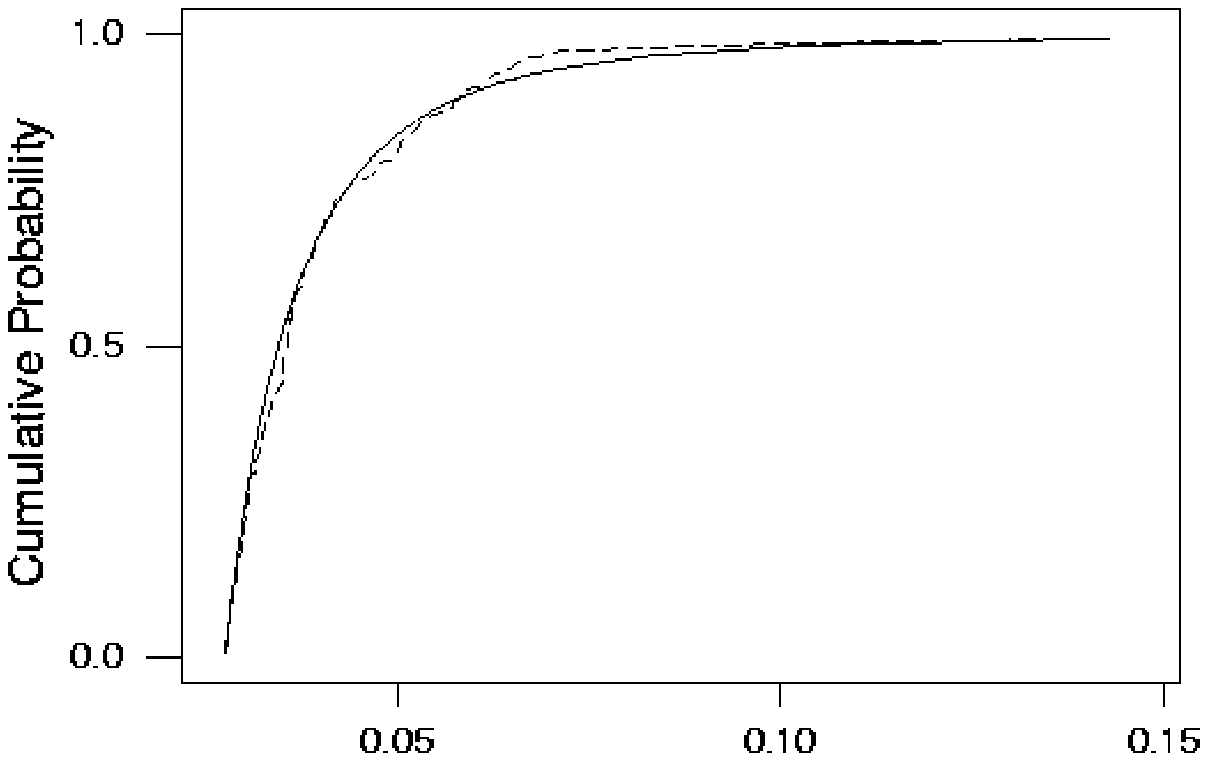}}
\caption{Empirical (dash) and Fitted (solid) CDF for differences $d^{*}_{i}$}
\label{plot2}
\end{figure}

\noindent
Following Gopikrishnan et al. ~\cite{GOPI98,GOPI99}, Liu et al. ~\cite{liu1}, 
and Mantegna and Stanley \cite{mantegna}, we wish to test statistically, the
null hypothesis that these extreme observations are consistent
with the Power-law distribution. 
\\
For the data set $d_{1},\ldots,d_{126}$, the estimates of the parameters are
$\tilde{\gamma}=0.0313$ and $\hat{\alpha}=3.31$; for which the values of the
test statistics were found to be $A^{2}=0.5778$ and $W^{2}=0.0995$.
\\
The second data set, consisting of the absolute values $d^{*}_{1},\ldots,d^{*}_{144}$
of the smallest differences, we find $\tilde{\gamma}=0.0275$, $\hat{\alpha}=3.05$;
the values of the test statistics were $A^{2}=0.7142$ and $W^{2}=0.1264$.
\\
Referring to the asymptotic percentage points in table 3, we find that, in both cases,
the $p-$value is greater than 0.25, indicating a very good fit to the power-law
distribution. Figures 1 and 2 show the fitted and empirical cumulative distribution
functions for both data sets.

\section{Conclusions}
\label{Conclu}

In this work, the use of the Anderson-Darling and Cram\'{e}r von-Mises goodness-of-fit statistics
have been presented for the case of the Power-law distribution and the asymptotic distributions of
these test statistics have been obtained. 
\\
Following \cite{DURBIN73}, the asymptotic distribution of statistics based on the empirical distribution
function does not depend on location and scale parameters, but it might depend on the value of a shape parameter;
here it has been shown that for the Power-law distribution, these asymptotic
distributions do not depend on the particular value of shape parameter $\alpha$ by calculation of
the covariance function of the corresponding limiting process on which they are based, relating it
to the case of the exponential distribution. When the threshold parameter $\gamma$ is known, simulations
results suggest that the asymptotic percentage points of the Anderson-Darling and
Cram\'{e}r von-Mises statistics can be used with good accuracy even for small n.
For the case of $\gamma$ unknown, an estimator was proposed. From Monte Carlo results, we
conclude that such an estimator is useful for goodness-of-fit purposes in the sense that it allows
the calculation of the test statistics, preserving the asymptotic distribution, although the speed
of convergence appears to be slower in this case.
\\
The proposed test was shown to be useful in analyzing stock price variations, where it is
required a significance test of fit for the power-law distribution. The results obtained
here for describing the changes in the Mexican stock exchange index IPC, are consistent
with previous studies where the power-law distribution with shape parameter $\alpha \simeq 3$ has been proposed.
\begin{acknowledgement}
A.R.H.M wishes to thank Conacyt-Mexico for financial support provided under Grant No. I35646-E/3399-E/5617. 
\end{acknowledgement}

\end{document}